**Anomalous magnetic behavior and complex magnetic structure of proximate $LaCrO_3$ – $LaFeO_3$ system**

Brajesh Tiwari[1*], Ambesh Dixit[2], M. S. Ramachandra Rao[3]
[1] Department of Physics, Institute of Infrastructure Technology Research and Management, Ahmedabad-380026, India.
[2]Department of Physics & Center for Solar Energy, Indian Institute of Technology Jodhpur, Karwad 342037, India.
[3]Department of Physics, Indian Institute of Technology Madras, Chennai, 600036, India.
*brajeshtiwari@iitram.ac.in

**Abstract:**

We investigated complex magnetic properties of multifunctional $LaCrO_3$-$LaFeO_3$ system. The magnetic measurements substantiate the presence of competing complex magnetic ordering against temperature, showing paramagnetic to ferrimagnetic transition at ~ 300 K, followed by antiferromagnetic (AFM) transition near ~250 K superimposed on ferrimagnetic phase. The onset of weak ferrimagnetic ordering is attributed to the competing complex interaction between two AFM $LaCrO_3$-$LaFeO_3$ sublattices. The low-temperature AFM ordering is also substantiated by temperature-dependent Raman measurements, where the intensity ratio of 700 $cm^{-1}$ Raman active mode showed the clear enhancement with lowering the temperature. The non-saturating nature of magnetic moments in $LaCrO_3$-$LaFeO_6$ suggests the predominating AFM ordering in conjunction with ferrimagnetic ordering between 250 K – 300 K up to 5 T magnetic field. A complex magnetic structure of $LaCrO_3$-$LaFeO_3$ is constructed, emphasizing the metastable magnetic phase near room temperature and low temperature antiferromagnetic state.

**Keywords:** Magnetism; proximity effect; ternary magnetic oxides, Raman Spectra, transition temperature; ferrmimagntism; antiferromagentism.

**Introduction:**

The multifunctional materials, especially complex oxide materials, are not only attracting attention due to their potential but also providing rich understanding of the fundamentals, which allows designing novel materials with desired functional properties. $LaTMO_3$ (TM = Transition Metal) is one such family of complex oxide materials, having perovskite structure with TM magnetic ions. Direct or indirect cross –coupling among spin, orbital, lattice and charge degrees of freedom provide avenue for potential application and fundamental study in these materials.  The canted spin structure of TM ions also exhibits competing magnetic interaction and thus, giving rise to the intricate magnetic structures. $LaCrO_3$ and $LaFeO_3$ oxide systems exhibit antiferromagnetic transitions at ~ 290 K and 740 K respectively [1-3, 4, 5] and more interestingly, also exhibit weak ferromagnetisms near room temperature [6-9]. The search of more than one ferroic ordering in oxide systems is always attracting attention for their potential in new class of electronic devices such as four state memories, voltage controlled magnetic switches and sensors, electric field controlled spintronic devices. The recent studies provide evidences about the room temperature magnetoelectric coupling in $LaCrO_3$ [6] and probably the ferroelectric ordering in $LaFeO_3$ perovskite systems [5]. In spite of magnetic and magnetodielectric properties, these oxide systems exhibit enhanced oxygen ionic conductivity and their electronic conductivity can be tailored by manipulating the suitable dopant at different cation sites. For example, the calcium doping at La site in $LaCrO_3$ makes it highly conducting and is a potential candidate for high temperature solid oxide fuel cell electrode material. The distorted $TMO_6$ octahedra in $LaTMO_3$ systems is the driving cause for the complex physical properties such as variation of transition temperature, strong electron-phonon coupling, weak ferromagnetism, electrical conductivity by manipulating the exchange and hopping strengths [10]–[12]. The order and amplitude of such changes in physical properties are associated with the degree of distortion. The divalent doped lanthanum manganite system shows charge –ordering, which is closely related to antiferromagnetic phase, while charge delocalization i.e. metallic state coincides with ferromagnetism. The screened potential energy becomes large in certain TM oxide materials, due to the various external factors (doping, temperature, pressure etc), causing electron localization and thus, inhibiting the electrical conduction. This is known as Mott – transition in such transition metal-based perovskite systems [13], [14]. In addition, the hole-doped $La_2CuO_4$ antiferromagnetic becomes high temperature superconductor because of the strong electron-electron correlation. These systems provide a wide avenue to understand the underlying physics and related mechanism governing such functional properties and their possible tunability. The double perovskite materials with and without transition metals are gaining attention for designing multifunctional material systems [15] [16]–[18]. Way back in 1960's, Goodenough-Kanamuri predicted that double perovskite $La_2CrFeO_6$ should have a ferromagnetic ground state with Tc close to room temperature [19]–[21]. Since then a lot more efforts are made experimentally and theoretically with varying conclusions. Some concluded ferromagnetic ground state [20] while others ferromagnetic [21]. This prompted us to explore $La_2CrFeO_6$ double perovskite while synthesizing $LaCrO_3$-$LaFeO_3$ system. This system is usually showing intermediate properties of both perovskites, however, some unique magnetic and optical phonon properties are also noticed due to close proximity of these two predominantly antiferromagnetic compounds. It is a fundamental challenge in materials' design to control and understand the change in materials behavior in close proximity to other materials under varying external conditions. In the present work, we investigated the effect of spin ordering on the magnetic and electronic properties of closely proximated $LaCrO_3$-$LaFeO_3$ system.

## Experimental Details:

All oxide precursors were heated at ~ 800 $^oC$ to remove any residual oxide and ground in a stoichiometric ratio to homogenize the pre-synthesized materials. This homogeneous material was heated at 950 $^oC$ for 48 hours with intermediate grinding to ensure the homogeneity of a solid solution. For structural and phase identification, powder X-ray diffraction (XRD) data of the samples were recorded using a PANalytical X'Pert Pro X-ray diffractometer with Cu Kα radiation. DC magnetic measurements were performed using a vibrating sample magnetometer as an attachment in Physical Property Measurement System (Model 6000, Quantum Design, USA) in the temperature range of 100–350 K. Magnetization measurements were performed as a function of temperature in zero-field cooled (ZFC) and field cooled (FC) modes. Various magnetization isotherms were recorded at different temperatures up to an applied magnetic field of 50 kOe in the vicinity of magnetic transitions. Raman spectra were recorded for $LaCrO_3$, $LaFeO_3$ and proximate $LaCrO_3$-$LaFeO_3$ at room temperature with the help of 532 nm green laser source. Temperature-dependent Raman spectra were also recorded for $LaCrO_3$-$LaFeO_3$ system down to 100 K in order to understand the spin-phonon coupling if any, following the magnetic behavior.

## Results and discussion:

The phase identification of synthesized materials was confirmed using X-ray diffraction and the respective diffractograms are shown in Fig. 1 in conjunction with $LaCrO_3$ and $LaFeO_3$ perovskite structure to understand the phase evolution of $LaCrO_3$-$LaFeO_3$ system. The XRD patterns, Fig. 1 (lower and middle panel), confirm the phase purity of pristine $LaCrO_3$ and $LaFeO_3$ bulk materials and results are consistent with the reported literature [6], [7], [22]. All the peaks are in agreement and representative (h k l) planes are marked for $LaCrO_3$ system. $LaCrO_3$ and $LaFeO_3$ systems crystallize in distorted orthorhombic perovskite system with almost similar lattice parameters. The structure consists of corner shared tilted $TMO_6$ (TM = Cr, Fe) octahedra. The structural and magnetic details as reported earlier suggest that the tilted octahedral may induce non-collinearity in the spin structure, giving rise to the weak ferromagnetism in these systems [2], [6]. The XRD diffractogram, Fig. 1 (upper panel), can be visualized as the superimposed XRD spectrum of these pristine materials. These closely spaced doublets confirm the formation of mixed phase $LaCrO_3$-$LaFeO_3$ system without any additional impurities.

The room temperature Raman spectrographs of these systems are shown in Fig. 2 and analyzed to understand the microscopic phase evaluation for $LaCrO_3$-$LaFeO_3$ system. The factor group analysis of orthorhombic (Pnma space group) suggests that there are 24 ($\Gamma = 7A_g + 5B_{1g} + 7B_{2g} + 5B_{3g}$) Raman active modes in this distorted perovskite $LaCrO_3$ and detail of mode assignments are given in references [6], [23], [24]. The identical crystallographic structure of $LaCrO_3$ and $LaFeO_3$, space group Pnma, gives rise to the similar vibrational modes for both systems with a small deviation for different atomic masses of Cr and Fe atoms. Thus, it was difficult to separate out the vibrational contribution of one from other, as evident in XRD graphs, Fig. 1, as well as from Raman spectra, Fig 2. The temperature dependent Raman spectra are shown in Fig 2, where some of the Raman active modes in proximate $LaCrO_3$-$LaFeO_3$ system show peculiar temperature dependence as compared to pristine systems. After careful analysis of first and second order optical phonon $B_{2g}(1)$ modes, it is observed that intensity ratio ($I_1/I_2$) of these modes show a sharp increase near second magnetic transition, as shown in Fig 2(b). This increase in the intensity of first order $B_{2g}(1)$ mode as compared to second order mode near magnetic anomaly indicates a spin-dependence of optical phonon mode in Raman scattering that can be a manifestation of electron transfer with lattice vibrations and/or anisotropic exchange interactions.

$LaCrO_3$ and $LaFeO_3$ materials are known antiferromagnetic with Neel temperature 290 K and 710 K, respectively [6], [9], [25]. The tilted Cr/Fe-$O_6$ octahedra lead to the canted TM electron spins and thus, causing weak ferromagnetism in these systems. In conjunction with the observed weak ferromagnetism in these systems, the near room temperature magnetodielectric coupling has also been reported in both systems. Considering the complex magnetic interactions in pristine systems, DC magnetization as function of temperature from 350 K to 100 K has been recorded under zero field cooled (ZFC) and field cooled (FC) condition at 1000 Oe for $LaCrO_3$-$LaFeO_3$ system. The measured temperature dependent magnetic moment is shown in Fig. 3. The observed sudden rise in magnetization near 290 K in this sample is due to the antiferromagnetic ordering of $LaCrO_3$, superimposed with weak ferromagnetism because of spin canting. The proximate presence of $LaFeO_3$ tries to reorient the magnetic spin structure of $LaCrO_3$ sublattice along the weak ferromagnetic structure of $LaFeO_3$, causing relatively larger ferromagnetic component below 290 K. This ferromagnetic state of $LaCrO_3$-$LaFeO_3$ system preserves down to 250 K, after that the antiferromagnetic ordering of $LaCrO_3$ starts dominating. This $LaCrO_3$ antiferromagnetic dominance, leads to another antiferromagnetic transition for $LaCrO_3$-$LaFeO_3$ system. Thus, two magnetic transitions are observed clearly at 290 K (weak ferromagnetic) and 250 K (antiferromagnetic) in $LaCrO_3$-$LaFeO_3$ system. Further, to understand the dynamic nature of these magnetic transitions, we carried out temperature dependent AC magnetic measurements at 100, 300, 1000, 3000 and 10000 Hz frequencies in the temperature same range and plots are shown in Fig. 3. The first magnetic transition around 290 K is coinciding to that of $LaCrO_3$ long-range antiferromagnetic Neel temperature, superimposed with weak ferromagnetic ordering of both pristine $LaCrO_3$ and $LaFeO_3$ systems [4]. The onset of additional magnetic transition at ~ 250 K may be the consequence of competing spin interaction between close proximity of magnetic Cr and Fe sublattices. To our surprise even transition at 250 K is also frequency independent, suggesting a long-range spin ordering in $LaCrO_3$-$LaFeO_3$ system. The observed frequency independence magnetic transitions also rule out the possibility of any clustering/spin glassy impurities in the synthesized $LaCrO_3$-$LaFeO_3$ system. The simultaneous presence of two AFM transitions in this system may be the consequence of proximity of Fe and Cr magnetic ions and complex magnetic interaction between them. The XRD results confirm the synthesis of $LaCrO_3$-$LaFeO_3$ mixed phase system and observed complex magnetic properties suggest the presence of competing magnetic interaction between different Cr and Fe ion sites in $LaCrO_3$-$LaFeO_3$ system. The magnetization isotherms are measured near these transition temperatures to probe the nature of magnetic ordering in conjunction with the measured temperature dependent magnetization measurements. The measured magnetic isotherms are shown in Fig. 4 for temperatures 200, 250, 280 and 315 K. The weak ferromagnetic component at 315 K is lower than that of 280 K, suggesting that at higher temperature, only $LaFeO_3$ weak ferromagnetic component is contributing in this system. However, at lower temperatures, the contribution of $LaCrO_3$ weak ferromagnetic component is also added, as can be observed in Fig. 4. In contrast to the observed weak ferromagnetism, the magnetization curves are not saturating up to 50 kOe magnetic field suggesting the dominance of antiferromagnetic state in $LaCrO_3$ – $LaFeO_3$ system. The magnified magnetization curves are shown in Fig. 4 (lower panel), showing nearly temperature insensitive weak remnance ~ $3x10^{-3}$ emu/g. However, the respective coercive field is much larger ~ 1 kOe for $LaCrO_3$ – $LaFeO_3$ system, suggesting the robust metastable ferromagnetic state.

The magnetic phase diagram against temperature is constructed for $LaCrO_3$-$LaFeO_3$ system, and is shown in Fig. 5. The high-temperature phase (>740 K) is paramagnetic, changing into antiferromagnetic phase dominated by $LaFeO_3$ in the proximate $LaCrO_3$-$LaFeO_3$ system. This antiferromagnetic phase is superimposed with the weak ferromagnetic phase because of the canted iron spins in $FeO_6$ octahedron in $LaFeO_3$ sublattice, as marked in Fig. 5. Further reducing the temperature below 290 K, this changes into a metastable ferromagnetic phase at the onset of $LaCrO_3$ antiferromagnetic transition. This phase persists until 250 K, where the system shows another antiferromagnetic phase with weak ferromagnetism simultaneously.

## Conclusion:

We studied the complex magnetic behavior of proximate $LaCrO_3$- $LaFeO_3$ system with different magnetic phases and intertwining of optical phonons with magnetic ordering. These studies may lead to the materials engineering to design complex magnetic structured materials with competing magnetic phases at or above room temperatures in mixed phase systems. The observed spin-lattice coupling from temperature dependent Raman spectra shows a possibility of inducing magnetodielectric coupling in such mixed phase systems. Further investigations are required to understand the microscopic origin of observed complex magnetic structure and spin-lattice coupling proximate $LaCrO_3$-$LaFeO_3$ system for possible tunability of spin-lattice functional properties.

## Acknowledgement:

Author Brajesh Tiwari acknowledges Professor Shiva Prasad for technical discussions for the manuscript. Ambesh Dixit acknowledges UGC-DAE Consortium For Scientific Research, Gov. of India through project number CRS-M-221 for this work.

**Figure Captions:**

1. X-ray diffraction of bulk $LaCrO_3$ and $LaFeO_3$ material with $LaCrO_3$ (Bottom panel) and $LaFeO_3$ (Middle panel).
2. (a) Raman Spectra of $LaCrO_3$, $LaFeO_3$ and $La_2FeCrO_6$ at room temperature recorded using 532 nm laser source. (b) Temperature-dependent Raman spectra of $La_2FeCrO_6$ using 514 nm laser source. (c) Intensity ratio of first order and second order Raman peak as a function of temperature suggesting suppression of second order peak upon magnetic ordering.
3. Temperature-dependent DC (Zero-Field Cooled and Field Cooled) and AC (at different frequencies) magnetic susceptibility plots for $La_2FeCrO_3$ .

4. Representative isothermal magnetization loops close to magnetic transitions.

**Figures:**

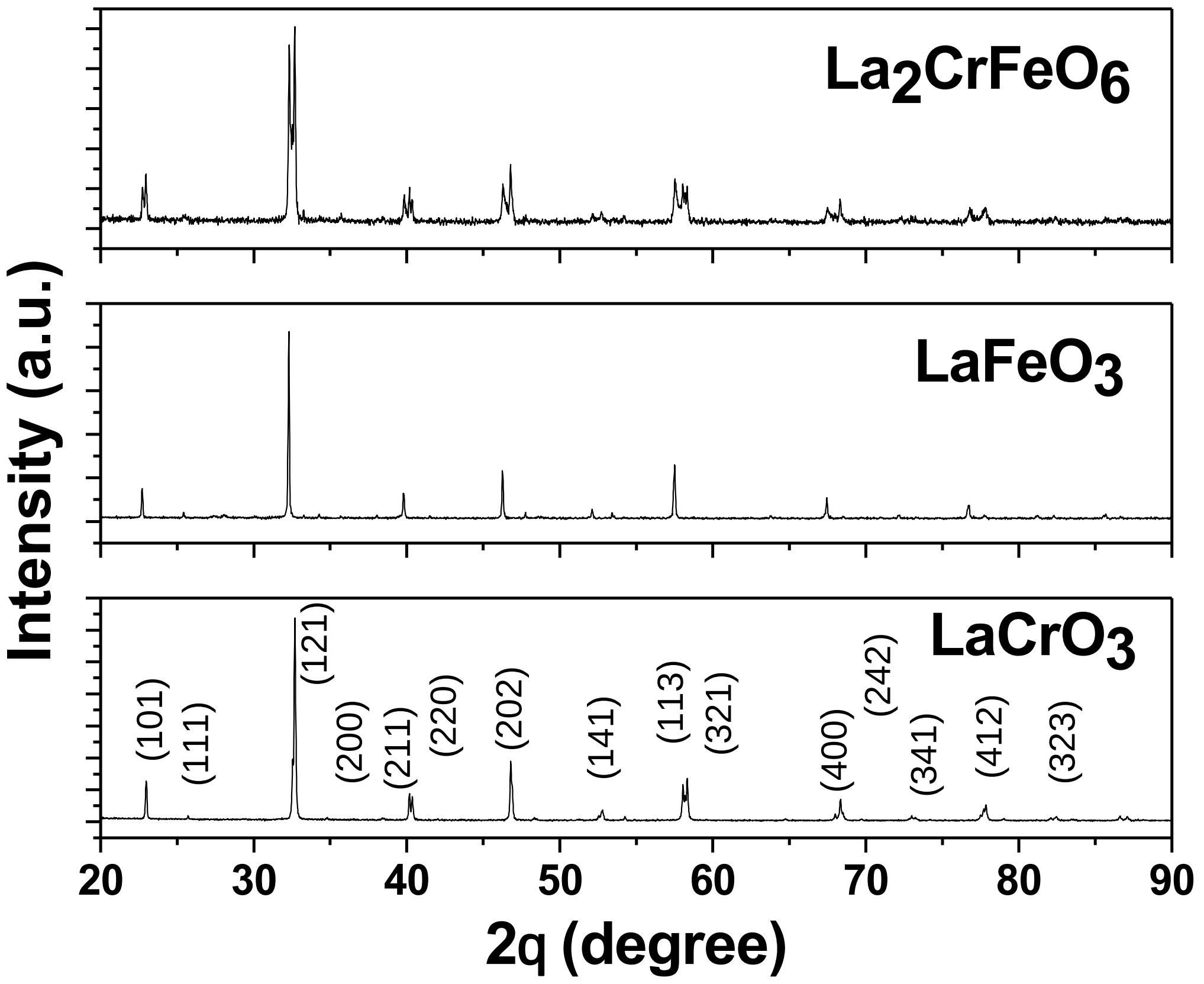

**Figure 1.**

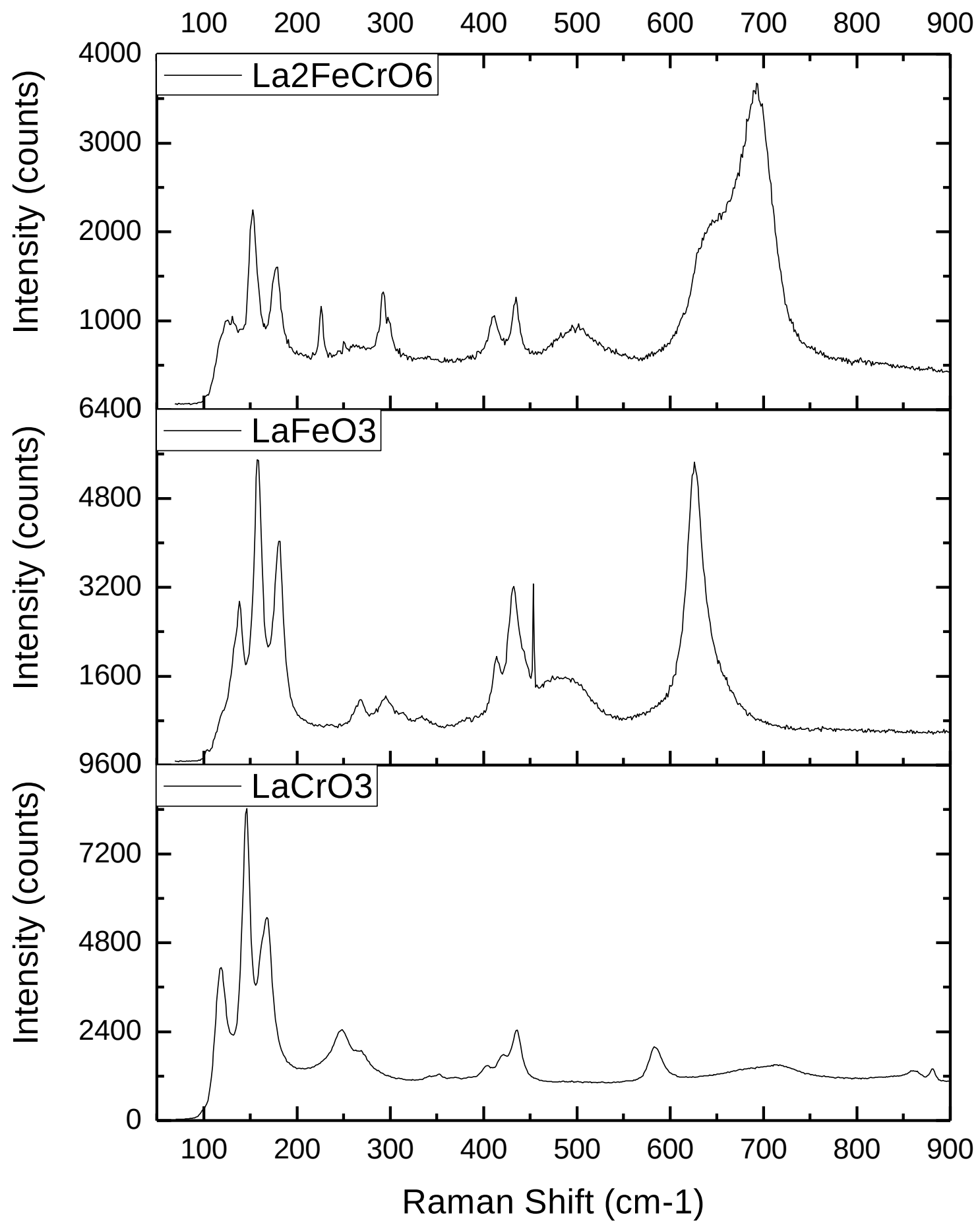
100 200 300 400 500 600 700 800 900
4000
3000
2000
1000
La2FeCrO6
Intensity (counts)
6400
4800
3200
1600
LaFeO3
Intensity (counts)
9600
7200
4800
2400
0
LaCrO3
Intensity (counts)
100 200 300 400 500 600 700 800 900
Raman Shift (cm-1)

**Figure 2(a)**

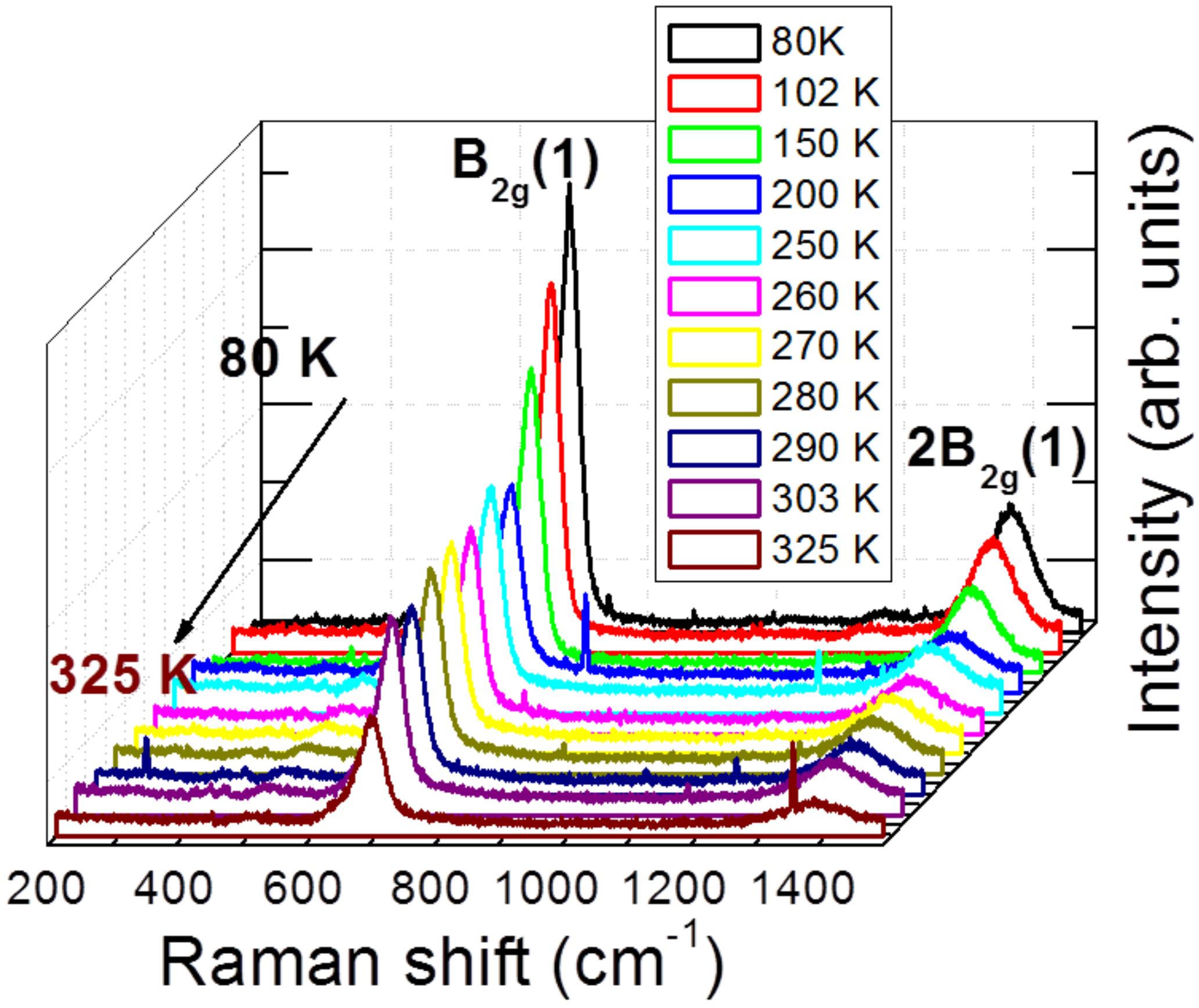

**Figure2 b**

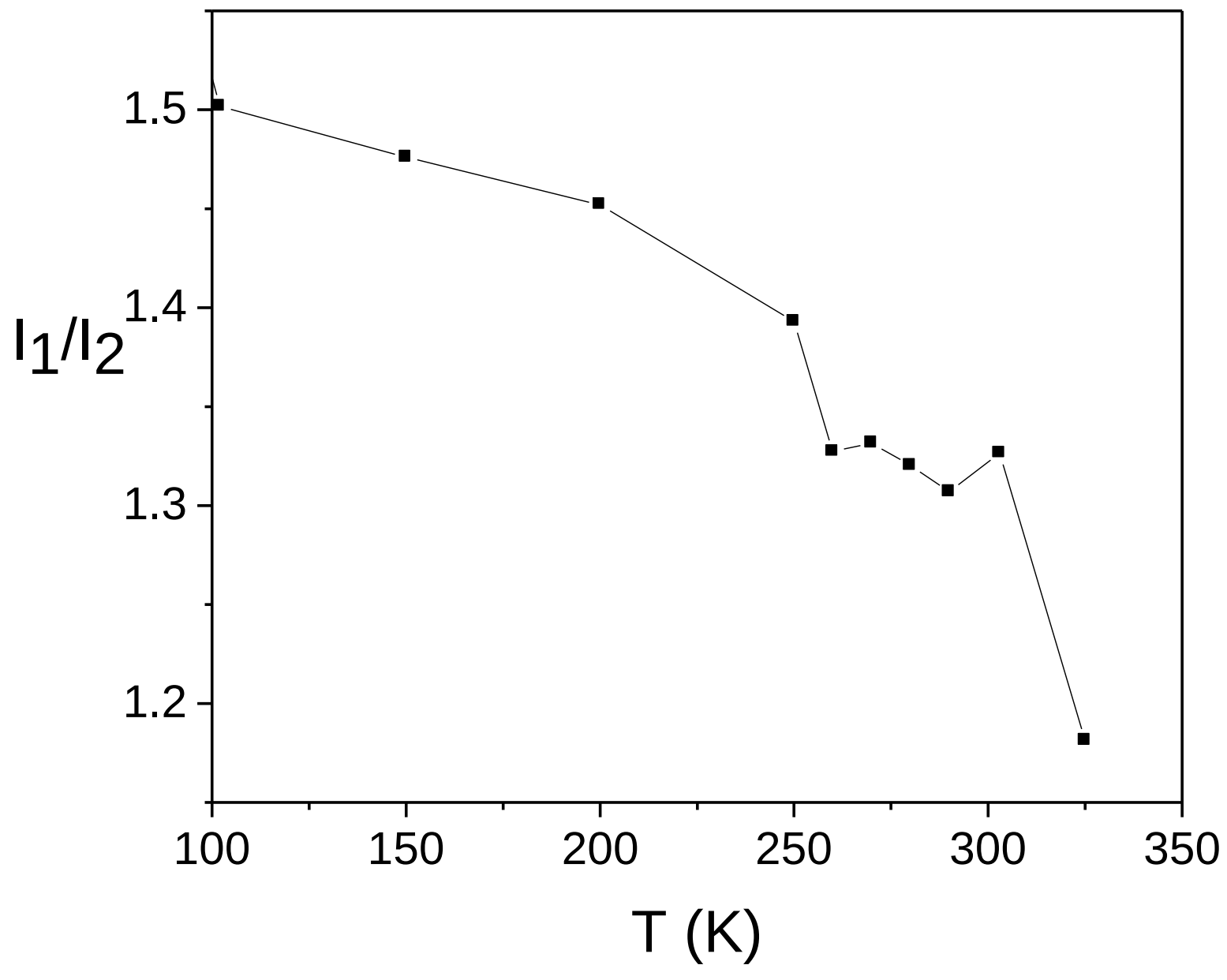

**Figure 2 (c)**

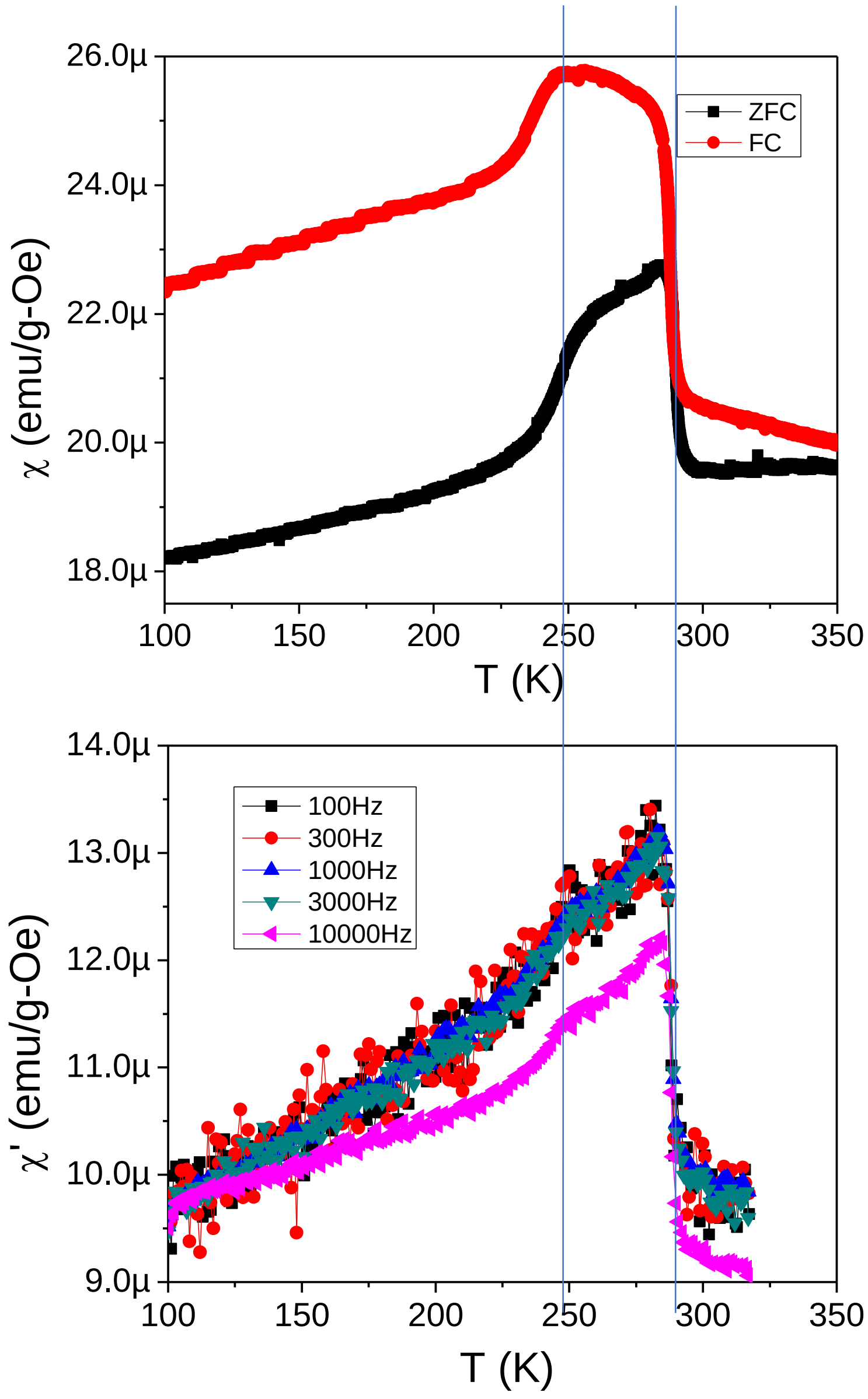

**Figure 3**

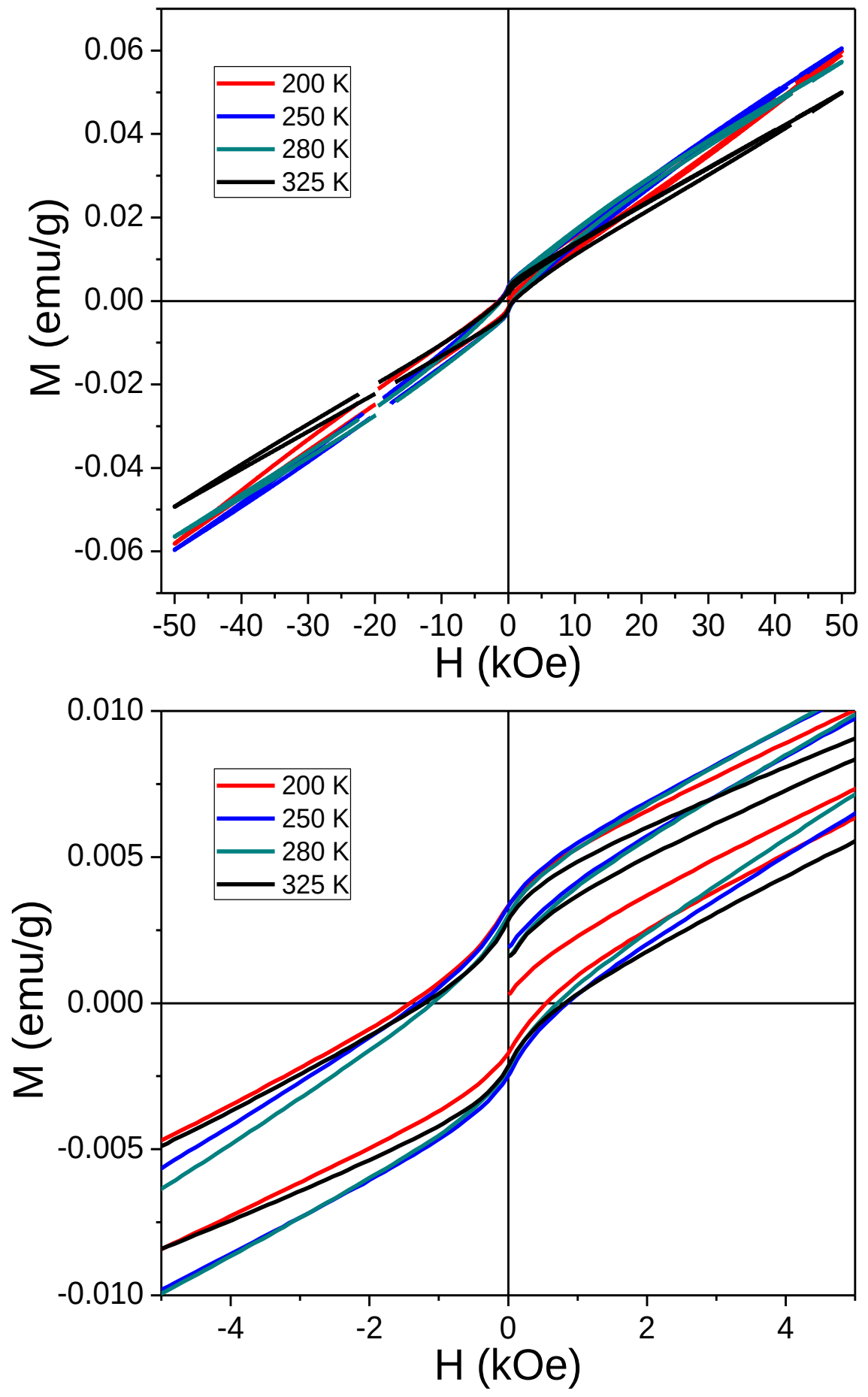

**Figure 4**

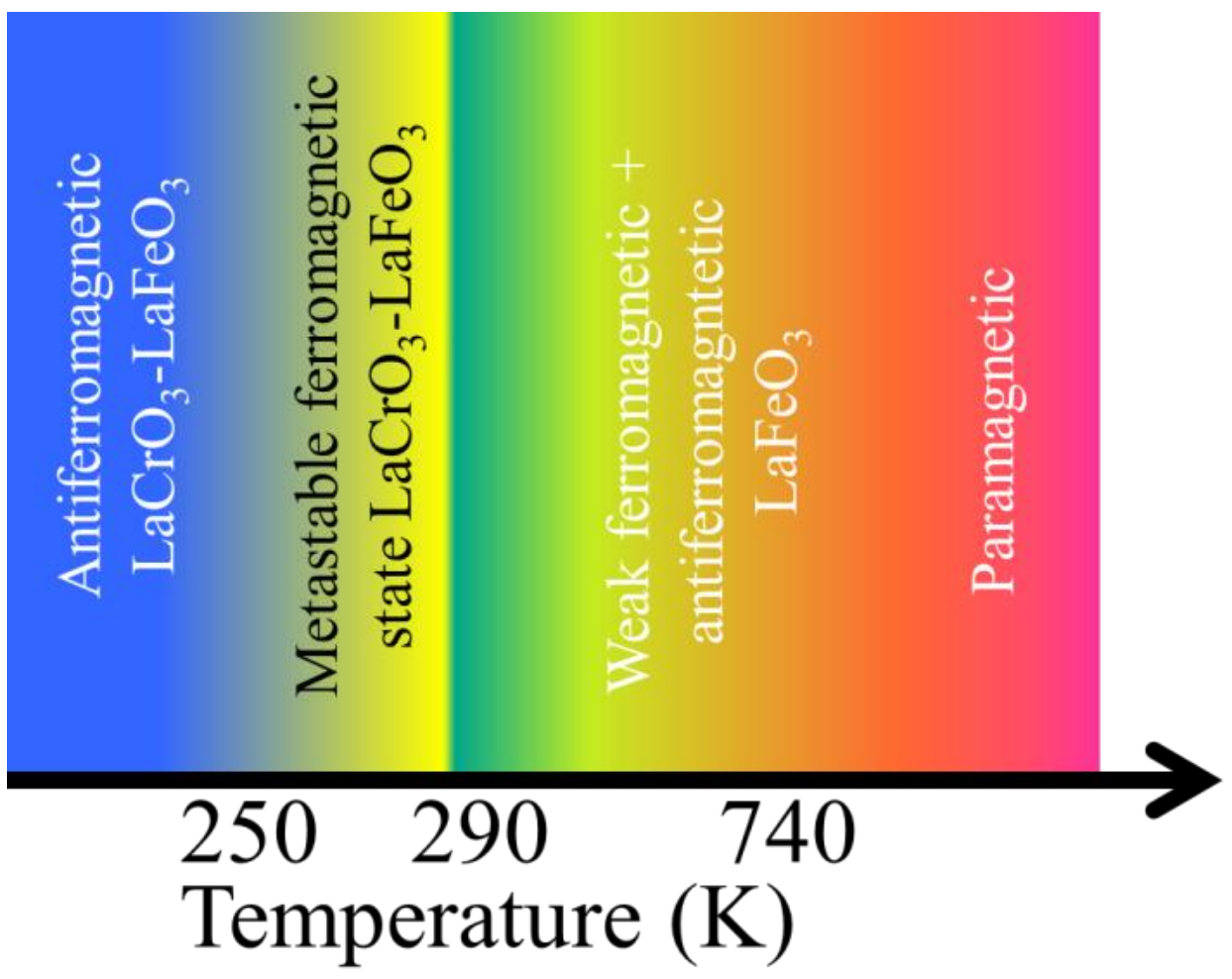

**Figure 5**